\documentclass[aip, rsi, amsmath, amssymb,reprint,]{revtex4-1}
\usepackage{graphicx}
\usepackage{dcolumn}
\usepackage{bm}
\usepackage[linktocpage=true]{hyperref}
\hypersetup{colorlinks=true,citecolor=red,urlcolor=blue}
\usepackage[utf8]{inputenc}
\usepackage[T1]{fontenc}
\usepackage{mathptmx}
\usepackage{etoolbox}
\usepackage[usenames,dvipsnames]{xcolor}
\usepackage{color,soul}
\makeatletter
\def\@email#1#2{
 \endgroup
 \patchcmd{\titleblock@produce}
  {\frontmatter@RRAPformat}
  {\frontmatter@RRAPformat{\produce@RRAP{*#1\href{mailto:#2}{#2}}}\frontmatter@RRAPformat}
  {}{}
}
\makeatother
\begin{document}
\preprint{}

\title{Tutorial: Current controllers for optimizing laser cooling on cold atom experiments}
\author{D. O. Sabulsky}
\altaffiliation{\textit{Please}, do not contact the author about these devices. Have a discussion with your local laboratory electronics support staff and/or open a copy of The Art of Electronics \cite{Horowitz:1981307}. \\ \\If you must, and/or for all other inquiries and comments: \\ \href{mailto:dylan.sabulsky@obspm.fr}{dylan.sabulsky@obspm.fr}}.
\affiliation{LNE-SYRTE, Observatoire de Paris–Universit\'{e} PSL, CNRS, Sorbonne Universit\'{e}, 61 Avenue de l’Observatoire, Paris F-75014, France}

\date{\today}

\begin{abstract}
The design of single chip current source based on a common power operational amplifier is presented and demonstrated for the purpose of controlling applied magnetic fields using bias/shim electromagnets in cold atom experiments. 
The efficacy of the design is realized via application to red-detuned polarization-gradient cooling of $^{87}$Rb down to 3 $\mu$K.
Further, we demonstrate Raman spectroscopy using these devices to apply current and so generate a precise, accurate, and reproducible magnetic field.
This work is intended as a short tutorial for new graduate students and postdocs of laser cooling and trapping. 
\end{abstract}

\maketitle

\par It is not uncommon to stumble upon what has become a familiar sight in both new and old cold atom laboratories - a collection of multi-channel bench-top power supplies, typically two channel 24 V / 3 A, often on the floor of a laboratory with water cooling present (!), directly supplying current to three pairs of quasi-Helmholtz electromagnets.
They are held together with zipties, vinyl tape, and/or epoxy; sometimes the number of turns contained within these electromagnets is known, often it is not, and coincidentally neither is the inductance, calculated or empirical.
While it is often the case that the atoms are sufficiently cold after sub-Doppler laser cooling to proceed to further intermediate cooling techniques or directly to optical or magnetic traps, the physical questions of optimizing the atom number and temperature, and so phase-space density, are often hastily dismissed as time consuming. 
This is a trade in time and efficiency for simplicity and poorly placed pragmatism, and bellies the fact that new students arrive with a new and different set of skills from their predecessors.
\par Computer controllable commercial current controllers are available, but a series of simple electronics circuits are capable of fulfilling this role from a single low noise power supply of the user's choice, and at a more affordable cost, with student education as an added benefit.
The design, construction, and testing of these electronics is a learning experience that new students of laser cooling and trapping should not dismiss - there is physics, electronics, and good systems design to be found in this exercise. 
\begin{figure} [h!]
\centering
\includegraphics[width=\linewidth]{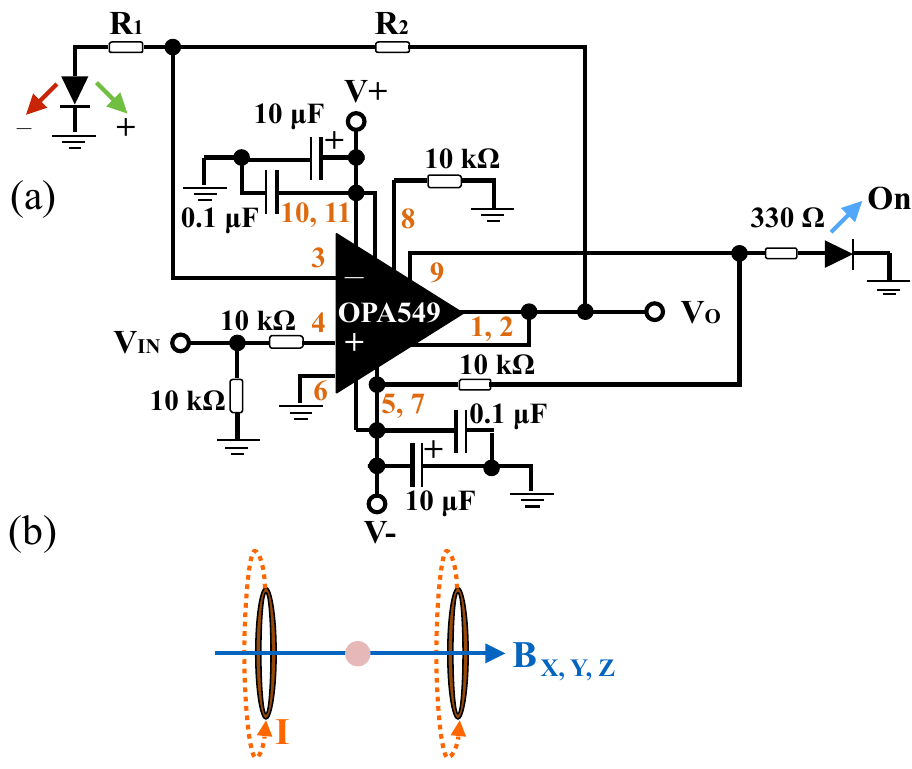}
\caption{An OPA549 as a single channel, bipolar, single chip current drivers for electromagnet pairs, a schematic \cite{Sabulsky2018,opa549}. 
(a) A Texas Instruments Burr-Brown OPA549 is the bipolar current supply - oriented as a non-inverting amplifier with closed-loop gain $G=1+\frac{R_{2}}{R_{1}}$, with a current limit resistor and LED indicators for when the driver is enabled and for the direction of the current (pin numbers shown in \textcolor{RedOrange}{\textbf{orange}}).
This circuit, powered by $\pm$12 V in constant voltage mode, is set with a 4 A maximum output current via the current limiting resistor, and gives an output that scales as 4.5 A/V with unity gain (\textbf{R}$_{1}$ = 1 k$\Omega$, \textbf{R}$_{2}=0$)\footnote{This is not universal, but is for the resistance of the given electromagnetic.} - given the control electronics (a 12-bit DAC), the current precision realized is 11 mA per voltage step of 2.4 mV\footnote{Limited by the aforementioned 12-bit DAC.}, which is sufficient current accuracy to realize magnetic fields allowing for sub-Doppler cooling near the recoil limit.
(b) This circuit drives a single matched pair of electromagnets for generating a magnetic field - typically in or near a Helmholtz configuration.
Reproduced and adapted with permission from D. Sabulsky, ``Constraining theories of modifed gravity with atom interferometry", Ph.D. dissertation (Imperial College London, 2018) \cite{Sabulsky2018}.
\label{im:1}} 
\end{figure}
\par Here, we focus on a key application within cold atom physics - controlling the background and stray magnetic fields for realizing one of the sub-Doppler cooling techniques commonly employed, red-detuned $\sigma^{+}-\sigma^{-}$ polarization gradient cooling \cite{PhysRevLett.61.169, Dalibard:89, Salomon_1990, Valentin1992,Wineland:92}.
A more modern example from laser cooling and trapping that requires current and therefore applied magnetic field precision would be gray molasses cooling \cite{Fernandes_2012,PhysRevA.87.063411,Salomon2013,Rosi2018}.
This tutorial is motivated by the  disparity in reported laser cooling results using the same atomic species and sub-Doppler technique as well as intransigence experienced in laboratories about the inadequacy of using bench-top power supplies directly.
Nominally, the aforementioned techniques should leave cooled ensembles at only a few recoil energies above the recoil limit, often the $\mu$K level for heavier atomic species. 
Practically, these temperatures are only obtained when stray magnetic fields that would modulate or shift the $m_{\text{F}}$ states relative to one another are carefully controlled; this has been pointed out succinctly and long ago in a common Undergraduate/Master's level texts \cite{Metcalf1999,Foot}, however, a brief summary of the known physics is called for.
\section*{Magnetic field control for optimal cooling}
\par Sub-Doppler cooling is extinguished when the loss of kinetic energy to the light field nears the recoil energy imparted by spontaneous emission: 
\begin{equation}
    U_{0} \simeq E_{r},  
\end{equation}
so, when a Zeeman shift is comparable to the scale of the light shift, $U_{0}\sim \mu_{\text{B}}B$, and assuming each degree of freedom has a kinetic energy term $\frac{1}{2} k_{\text{B}}T_{r}$, we can write
\begin{equation}\label{criterion}
    \mu_{\text{B}}B \simeq \frac{1}{2} k_{\text{B}}T_{r}.
\end{equation}
At this point there is no net energy loss in the optical pumping between $m_{\text{F}}$ levels, such that the lowest achieved temperatures with these techniques are a few times the recoil energy, the so-called recoil limit \cite{Steck03}: 
\begin{equation}
    k_{\text{B}}T_{r} = \frac{h^{2}}{m \lambda^{2}}.
\end{equation}
The presence of the Zeeman shift is one of the limits of this laser cooling technique - to maximize cooling, stray magnetic fields need to be controlled to keep the Zeeman sublevels degenerate.
\par Eq. \ref{criterion} is clear - for example, an ensemble of $^{87}$Rb cooled to 5 $\mu$K implies that $B< 7.7 \times 10^{-5}$ T, so 77 $\mu$T \footnote{see section 8.6 and equation (8.10) in Metcalf and van der Straten\cite{Metcalf1999} for the details of the calculation (there is a small factor to contend with). See Dalibard and Cohen-Tannoudji\cite{Dalibard:89} for the most details}. %of 0.097 
For scale, this is an order of magnitude less than the Earth's own magnetic field ($\sim 5 \times 10^{-5}$ T) in the laboratory!
This magnitude of magnetic field is not daunting to realize or to establish control over - a back-of-the-envelope calculation can help here; what is required to cancel Earth's ambient field in the laboratory? 
In the Helmholtz configuration, the magnetic field produced by a running current $I$ is $B_{\text{Helm}}=\left(\frac{4}{5}\right)^{3/2} \frac{\mu_{0} n I}{R}$. 
With modest $n = 50$-turn electromagnets of radius $R = 10$ cm, to generate a local field to compete with Earth's field of 77 $\mu$T requires about $172$ mA (assuming the axis of the two magnets is parallel to that of Earth's field). 
In addition to Earth's magnetic field, stray magnetic fields can be introduced by magnetic optomechanics or screws in the vacuum chamber, permanent magnets from nearby ion pumps, Zeeman slowers, or 2D MOTs, or even nearby experiments \footnote{Prior to 2014, an atom chip BEC experiment at Imperial College London was regularly disturbed by a superconducting magnet two floors above the experiment - in the years it took to locate and solve this 'problem', they installed active field compensation with circuitry similar to that presented. Lesson; know what is happening around your experiment.}.
Stray magnetic field from these devices lead to typical corrections range from around 10 $\mu$T to 200 $\mu$T, implying currents up to about 400 mA for electromagnets in or near Helmholtz configuration - higher current will generally be required for pairs of electromagnets falling well outside the Helmholtz conditions.
\par For an atomic ensemble of a few $10^{8}$ atoms generated with 25 mm optics and typical magneto-optical trap parameters, and a typical vacuum system, there is typically a relatively wide range of applied magnetic field values that allow for a sufficient state degeneracy in red-detuned polarization gradient cooling \footnote{This can be observed in (a) of Fig. 2. A literature review of thesis works showing plots like the aforementioned shows that many systems demonstrate a relatively insensitivity around the zero-field point.}.
This is in stark contrast to techniques like grey molasses that require a high field accuracy to realize optimal cooling. 
Typical bench-top power supplies offer LED displays elucidating the current applied down to the 10 mA or even 1 mA level - most boast analog control knobs, but some offer digital control.  
It is in this way that many laboratories are able to establish some level of control over stray magnetic fields to obtain cooling results $<25$ $\mu$K but often bottoming out near 10 $\mu$K. 
The electromagnet radius being too small, non-Helmholtz configurations, the number of turns versus the current fine tuning capability of the current supply, and other banalities of experiment can easily lead the control knobs of such a bench-top current supply to be insufficient for establishing sufficient magnetic field resolution zero-field conditions to complete sub-Doppler cooling at the $\mu$K level.
\par It is happy coincidence that bench-top supplies are sometimes sufficient as a current supply for bias/shim electromagnets in laboratories, despite few students doing the calculation to realize how often their experiments are actually amenable to this shortcut - or not \footnote{It is highly dependent on the electromagnet parameters, as most bench-top supplies are of similar characteristics. In the author's explorations, it was found roughly 30\% of bias electromagnets were being run sub-optimally with the current precision available from bench-top supplies.}.
Temperatures ranging from 1 to 3 $\mu$K employing red-detuned $\sigma^{+}-\sigma^{-}$ polarization gradient cooling on the $^{87}$Rb D2 line have been reported \cite{Mnoret2018,Sabulsky2020}.
These results are partially facilitated by the vast series of improvements to laser cooling systems and architectures under the auspice of the development of mobile quantum technologies.
These systems employ automated magnetic field control to maintain magnetic field stability, ensuring the continuity of the cooling results throughout the lifetime and location of the instrument.
We present a version of this type of circuit here, based on the results of recent theses that facilitated modern atom chip and atom interferometer experiments \cite{Sewell,Sewell2010,PhysRevLett.105.243003,Stammers,Cheng,Sabulsky2019,Sabulsky2020,Silva,Beaufils2022,junca:tel-03669058,Zou2022,Sabulsky2022}. 
\section*{OPA549 - a controllable current source}
\par We present an integrated solution based upon a single low-cost high voltage and high current bipolar operational amplifier for controlling bias/shim electromagnets on a cold atom experiment, originally adapted for driving current in atom chip wires \cite{Sewell,PhysRevLett.105.243003} and further modified for driving bias electromagnets for Raman spectroscopy and laser cooling \cite{Sabulsky2018,PhysRevLett.123.061102}.
The Texas Instrument's (TI) Burr-Brown products OPA549\cite{opa549}, functioning as single channel current sources, are used in the results presented here - see (a) of Fig.~\ref{im:1} for a schematic used in experiments.   
This power operational amplifiers can continuously output 8 A, with modest heat sinking \footnote{They should be thermally coupled with thermal paste to their metal laboratory enclosure or with chip-scale fins and some air-flow inside their box - please remember to check the correct grounding of these heatsinks. Fans are not strictly necessary. }. 
They are amenable to a parallel output configuration for increased controllable current output\cite{opa549} and can offer voltages amenable to driving electromagnets with significant resistance, $\sim\pm4$ V at 8 A directly from the chip.
The chip is protected against over temperature conditions (operating from -40 $^{\circ}$C to +125 $^{\circ}$C) and current overloads, as well as providing simple logic for disabling the output should these conditions be met or by other set conditions.
There is a simple connection for a user-selected current limit (critical for electromagnets!), as well as for thermal shutdown indication and/or forced shutdown - this chip, as a current driver, does not require a power resistor in series with the output current path, instead sensing the load indirectly. 
The chip contains an simple resistor controlled current limit functionality - a resistor or potentiometer connected between pins 6 and 8 will suffice, or connect them to realize the full current of the device. 
A small offset in the current limit circuit may introduce a variation of $\pm0.25$A for small currents - low output current applications may be better suited to other members of the family like the OPA547 or OPA548.
\par Control systems offering bipolar or unipolar voltage control, from 1 V to 10 V, can easily integrate with this current driver with some basic electronics before the non-inverting input to realize the full range of the device; one can even design further safety limits by limiting the control voltage \footnote{It is worth noting to the uninitiated - current driving is serious business and can damage and/or destroy critical components. Safety should be built into every step - within the computer/software level, at the voltage control, with the current limit, with the current output, even a fuse after the output isn't unreasonable pending the sensitivity of the application. Safety first!}.
While such a device can be employed in an inverting configuration, typically such driving power electronics are non-inverting to allow for active filtering to suppress high frequency noise and a reduction in loading effects on the source (i.e. the chip), a common find in specialty commercial current supplies. 
The gain of the non-inverting amplifier and voltage control input are both easily scaled, so control system agnostic. 
An optocoupler between any control system and current driving electronics is always strongly recommended, especially when using FETs and MOSFETs - this said, the author has deigned to use this circuit without such precautions, typically, and has found little cause for concern. 
\par It is common to add in a TTL control for fast switching of the circuit - an example implementation can be found in Zou's thesis, Fig. B.4 on pg. 128 \cite{Zou2022}.
This circuit is amenable to inclusion into a PID loop, either for cancelling ambient fields in conjunction with a Hall sensor or even just to feedback and stabilize the output current in an application - an example can be found in Sewell's thesis, Fig. 3.10 on pg. 55 \cite{Sewell} - wherein current stabilization comes from feedback via a sense resistor, in turn fed into a differential amplifier for comparison to the control voltage. 
Various indicators, like output enable, thermal shutdown, current direction, among others are possible if not encouraged. 
The current noise density (1 pA/$\sqrt{\text{Hz}}$ at 1 kHz) is far below what is required of the current driver for bias electromagnets and is sufficient for precision applications like current carrying wires in atom chips and for electromagnets driving magnetic fields required for precision spectroscopy. 
This chip has a gain bandwidth product of 900 kHz - for rf modulation, one might consider a different family of chips \footnote{When driving rf magnetic fields for atomic physics, never forget a transformer just before the application, they are usually superior to a DC block!}, but for driving bias magnets this is sufficient, even for rapid turning of magnetic fields in typical spin-polarized quantum gas experiments.
The gain-bandwidth decreases in an approximately linear relationship to 870 kHz at 125 $^{\circ}$C as the chip heats - it is mostly flat over the usable temperature range before thermal shutdown is enabled. 
\par For current supplies, one needs consider the current stability as a function of temperature.
Of note are the aforementioned current noise density as well as the thermal resistance of the chip - 1.4 $^{\circ}$C/W with an adequate heatsink \cite{opa549} or 30 $^{\circ}$C/W without any thermal sink \footnote{The author strongly recommends heat sinking these chips, especially the smaller cousins OPA547 and OPA548 - even at modest current, with insufficient heatsinks, the circuits oscillate between thermal shutdown and heating. While the chips are robust,it is quite annoying. Standard aluminium heatsinks with fins or even attaching the chip to the interior of their metal enclosure (that should be around all circuits... namely with a skin depth that protects against microwaves and rf, or just a few mm of material) is sufficient - follow correct and safe grounding instructions and protocols.}. 
Under maximum current output, the current limit as a function of temperature is flat - at the maximum, there is a slight non-linearity on approach to the maximum operation temperature. 
The input bias current also suffers a decrease, changing over the operational temperature range but is somewhat flatter near room temperature.
Further, for a given voltage powering circuit, the quiescent current suffers a quasi-linear decrease from room temperature to the maximum operating temperature. 
It is important the chip be well sunk to reduce these effects, despite the excellent qualities of the chip - good operating procedure is also important, waiting to take data until the chip is thermalized.
Last, with the output disabled, there is still a typical leakage current around $\pm$200 $\mu$A - an attached transistor, triggered by the output disable, can ensure this current doesn't leak to the application.  
\par Other chip families and manufacturers can be made into similar operational amplifier based current supply circuits of varying characteristics, without having to resort directly to FETs and/or MOSFETs. 
From TI, for example, the OPA family includes the likes of the OPA541, OPA544, OPA547, and OPA548 or even friends of the family like the LM1876 and related.
In the atomic sources of the MIGA infrastructure \cite{Sabulsky2020,junca:tel-03669058,Zou2022,Sabulsky2022,Beaufils2022}, the OPA547 \cite{opa547} is used, two in parallel for each channel (the OPA 548 would suffice as well, as a single chip) \footnote{A circuit diagram for the OPA547 as a current driver can be found in their datasheet \cite{opa547} or in Fig. B.4, pg 128 of a thesis from the MIGA project \cite{Zou2022}}.
%.
The LM1876 can be employed in pulsed operation atomic experiments that require reduced power consumption - when the amplifiers are idle, they switch to a power conservation mode with external logic.
More expensive options come from the likes of the PA02, PA16, and PA75 and other high current chips from Apex Microtechology \footnote{Evaluation boards are available, incredibly useful to start with for such expensive and high performance chips.} - while these devices would be underutilized for bias electromagnets given their cost and capabilities, such high current operational amplifiers can be used to drive precision inductive loads with a similar circuit to that presented here, and without a sense resistor. 
Be mindful of their current noise level, power dissipation requirements, and bandwidth with respect to the applied application.
Further, the majority of these chips are designed to drive current across resistive loads, not necessarily inductive loads - be prepared to create a RC snubber for your inductive load applications. 
These circuits are easily incorporated into existing experiments with a few extra analog voltage outputs to control the setpoint and a few TTL lines for fast switching of the circuit and so the electromagnets\cite{Dedman2001}, if required - they are perfect for PID loops and are readily integrated with Hall effect sensors for complete active feedback of the ambient magnetic field around an experiment.
\section*{Laser cooling}
\begin{figure*}
\centering
\includegraphics[width=\linewidth]{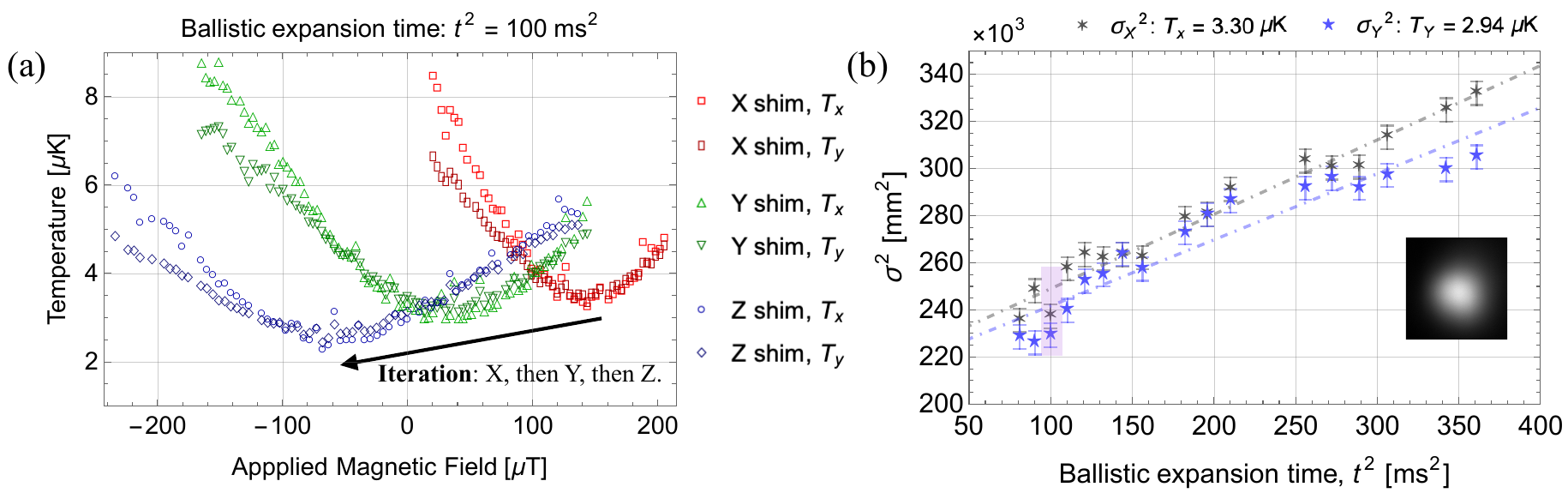}%FIX TEMP in (B)!!!
\caption{Applying control over the ambient magnetic field for red-detuned polarization gradient cooling. 
(a) Temperature of the ensemble as a function of the applied magnetic field. 
Starting with a scan of the applied magnetic field in the plane perpendicular to the axis of the 3D MOT electromagnets, X and then Y shim, followed by the magnet pair parallel with gravity, Z shim, the minimum temperature of the ensemble is found in a few iterations. 
While the electromagnets employed here were far from Helmholtz configuration, they produced a relatively flat field at the center but required current that could easily be given by the OPA549, but not by family members OPA547 or OPA548 without using a parallel output configuration. 
These data were taken with a fixed expansion time of 10 ms and each point is an average of five. 
(b) Time-of-flight temperature measurement realizing a 3 $\mu$K ensemble. 
Shading around 100 ms$^{2}$ indicates the data shown in (a).
The co-linearity of the data implies a mostly spherical ensemble, which is evident from a typical image of the ensemble, see inset, taken after 10 ms expansion time.
Time of flight measurements are typically less noisy - in the system exhibited here, the offset beatnote lock \cite{Ritt2004} of the D2 cooling laser was generally mismatched with the rapidity of this cooling sequence.
Each data point is an average of five. 
Part (a) is reproduced and adapted with permission from D. Sabulsky, ``Constraining theories of modifed gravity with atom interferometry", Ph.D. dissertation (Imperial College London, 2018) \cite{Sabulsky2018}.
\label{im:2}} 
\end{figure*}
\par To demonstrate these circuits in action, we present a $^{87}$Rb ensemble undergoing red-detuned polarization gradient cooling - the circuit shown in Fig.~\ref{im:1} formed the basis for a 2U 19" rack unit (depth 10 cm) with four OPA549 inside, three for each axis of compensation and an extra axis for bias in an atom interferometer experiment \cite{Sabulsky2018,Sabulsky2019}. 
The sequence, designed to generate cold atoms rapidly (total sub-Doppler cooling sequence $<$5 ms, instead of a more typical 10 to 25 ms seen on BEC machines) but without optimizing the phase-space density, is as follows: a 3D MOT is loaded from a 2D MOT with a push beam; the 3D MOT electromagnets are switched off; after a dwell time of 3 ms to allow eddy currents to subside \footnote{Each 3D MOT electromagnet is about 40 $\mu$H, and while their circuit stops current flow rapidly (a few $\mu$s!), the ringdown from the Eddy currents persists for 1 to 2 ms and is entirely expected and normal. Be sure to measure and characterize this is your own system with a pickup electromagnet and current clamp.}, the cooling begins.
This involves a linear sweep of the detuning from $-2.5\Gamma$ to $-25\Gamma$ \footnote{Any further, so much past the $^{87}$Rb F = 2 to F'= 2x3 crossover, realizes heating or at least an arresting of the cooling process as one nears the F = 2 to F' = 2 transition.} in 500 $\mu$s followed immediately by a linear sweep of the light intensity from a few $I_{\text{sat}}$ to zero in 500 $\mu$s \footnote{Please note,your mileage may vary, especially for such a rapid sequence of events - most systems are far more amenable to the more traditional method of red-detuned $\sigma^{+}-\sigma^{-}$ polarization gradient cooling wherein, without linear ramps, the laser frequency is stepped to around $-15\Gamma$ for between 10 and 20 ms and, concurrently, the laser intensity is stepped to a few I$_{\text{sat}}$ or less. This also produces the highest phase space densities realized with this technique}.
To exhibit the optimization the temperature of the ensemble via the applied magnetic field, an iteration is performed over scans of the applied magnetic field from three pairs of shim electromagnets, starting with the X shim magnet, see (a) of Fig.~\ref{im:2}, with a fixed time-of-flight of 10 ms and a fixed cooling sequence - note, the MOT electromagnets are oriented along the Z axis while XY are perpendicular. 
Fluorescence detection illuminates a CCD camera (Pike 505b), and 2D Gaussian fit routine, written in python, extracts the $\sigma_{x}$ and $\sigma_{y}$ \footnote{In the plane of the camera, not of the experiment! In this case, the camera is imaging 22.5$^{\circ}$ of the X axis in the XY plane and 16.5$^{\circ}$ down into Z; the atom widths measured are projections of all three primary axes} of the ensemble. 
\par Typically complete after two iterations, to ascertain the temperature a full time-of-flight expansion of the ensemble is then performed, (b) of Fig.~\ref{im:2}, realizing a minimum temperature of 3 $\mu$K \footnote{This temperature was actually too cold for the experiment \cite{Sabulsky2018,Sabulsky2019}, for various technical reasons, and so cooling was arrested at 5 $\mu$K!}, and without magnetic shielding.
This experiment would collect a few $10^{9}$ atoms in the 3D MOT, a few $10^{8}$ after the magnets are turned off, and end up with a few $10^{6}$ atoms at 3 $\mu$K.
Lower temperatures with similar atom numbers have been reported; they are realized due to improvements in specific implementations of the technique for the atomic/laser/control system \cite{Sabulsky2020,Beaufils2022,Sabulsky2022}.
\section*{Precision spectroscopy}
\begin{figure*}
\centering
\includegraphics[width=\linewidth]{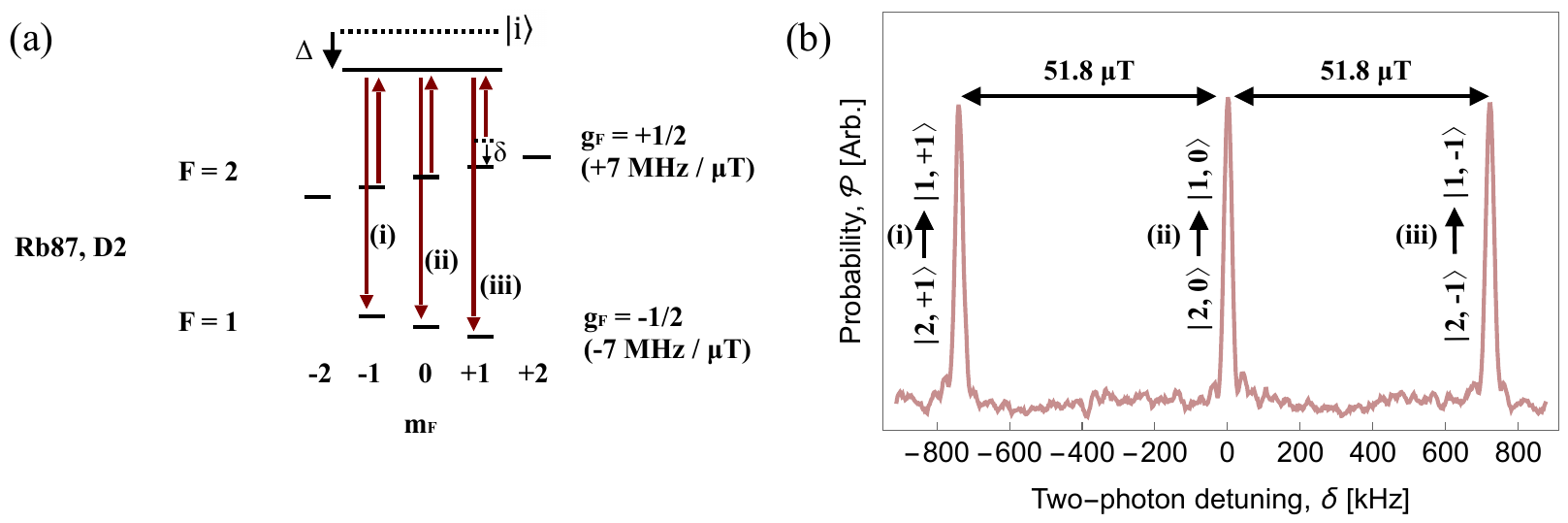} %Delta is one-photon detuning... delta is the two photon detuning.... YOU GOON FIX THAT. 
\caption{Co-propagating two-photon Raman spectroscopy with a 30 $\mu$s $\pi$-pulse.   
(a) Schematic level diagram for two-photon Raman transitions within the $^{87}$Rb ground states. 
The two-photon detuning $\Delta$ from a so-called imaginary state (merely highly red-detuned from the 5P$_{3/2}$ manifold) 
(i), (ii), (iii)
(b) Co-propagating Raman spectroscopy. 
Of note, no scalar, vector (in particular), or tensor light shifts are considered here, so one can then determine the magnetic field applied with confidence, finding an applied magnetic field of 51.8 $\mu$T, well within the linear low-field Zeeman splitting limit and lifting the degeneracy of the states within the F=1 manifold.
The Gaussian waist of the beam applying the spectroscopy pulse is twenty times larger than the Gaussian width of the atomic ensemble - the center of the beam is overlapped with the center of the ensemble. 
Each data point is an average of five. 
\label{im:3}} 
\end{figure*}
\par In addition to a role in sub-Doppler cooling techniques, this circuit is capable of sufficiently stable and precise current output for precision laser spectroscopy requiring applied magnetic fields, see Fig.~\ref{im:3} - they are otherwise amenable to current stabilization using a sense resistor \cite{Sewell}.
To explore this application, a 3D MOT of $^{87}$Rb is loaded from a 2D MOT with a push beam to realize a loading rate in excess of $\sim10^{9}$ atoms/sec. 
The atoms are subjected to red-detuned $\sigma^{+}-\sigma^{-}$ polarization gradient cooling applied over the six 3D MOT beams, reaching 3 $\mu$K. 
The atoms undergo a simple optical pumping routine before being probed via spectroscopy, wherein after polarization gradient cooling they are assured to be in the F = 2 manifold (equally distributed due to no magnetic field gradients\footnote{Two observations can tell you there is a magnetic field gradient; the magnetically sensitive transition will be inhomogeneously broadened to be larger than the applied Rabi frequency (compare to the magnetically insensitive transition), and the peaks will not be of equal probability $P$, instead showing a scaling where one of the magnetic states has the highest probability, the other state the lowest, with the insensitive state in between - it will look linear.}) by a precautionary illumination of the dark state for 100 $\mu$s. 
An applied magnetic field is stepped on along a direction of 45$^{\circ}$ from X or Y but within the XY plane.
A two-photon Raman pulse is applied to the ensemble; the two-photon detuning $\delta$ is scanned to explore the occupation density of the Zeeman sublevels via Raman spectroscopy.
There is no appreciable time-of-flight expansion for this test. 
\par A schematic diagram of the various allowed transitions using two-photon Raman transitions \cite{Foot, 1997}, see (a) of Fig.~\ref{im:3}, between the two hyperfine ground states of $^{87}$Rb are shown, and in the linear, or in the low field limit (i.e. small applied magnetic field) Zeeman splitting regime.
These internal energy level transitions, (i), (ii), and (iii) in Fig.~\ref{im:3}, are Doppler-insensitive co-propagating transitions, here shown from the F = 2 to the F = 1 ground states. 
The applied magnetic field can be determined with high precision, see (b) of Fig.~\ref{im:3}, if all the Stark effects i.e. light shifts \footnote{For example; the one-photon light shift, the two-photon light shift, the the intensity balance to minimize the aforementioned light shifts... and of course, we are concerned about scalar, vectorial, and tensorial components of the light shift which can all spoil our measurement of the magnetic field, among other things (see Fig. 4.4 in the relevant thesis \cite{Sabulsky2018} to visualize how)!} are controlled. 
\par With all precautions taken and the system calibrated, including good operating protocol of the circuits such that they are at their steady operational temperature - a few shots is usually sufficient, with good heat sinks - one can then determine the applied magnetic field at the position of the ensemble with high precision and accuracy. 
The measurement precision is limited by the frequency uncertainty of the laser system (for Raman systems used in atom interferometry, easily Hz level) and is certainly better than the current stability of the OPA549 current driver, while the measurement accuracy is limited the frequency chain and metrology of the laser system - this is certainly far superior to the current stability of the OPA549.
In (b) of Fig.~\ref{im:3}, the values are rounded to the hundred nT level \footnote{The systematic errors from spectroscopy are at the pT level, well below the current resolution and current noise of the circuit, converted to applied magnetic field.} which is more than sufficient to determine the stability of the applied field - in this experiment, we observe an applied field of 51.8 $\mu$T. 
\par The stability of the circuit was put to the ultimate test in a deterministic run pattern \cite{Hudson2014}, essentially turning an atom interferometer experiment into a lock-in amplifier \cite{Sabulsky2018, Sabulsky2019}. 
In 36 12-hour datasets, a modulation pattern included the change in the direction of the applied current to an electromagnet pair providing an applied magnetic field for driving counter-propagating Raman transitions. 
The modulation pattern, when demodulated and analyzed for contributions from errors in the magnetic field, revealed two results - one, and obviously, that the modulation pattern allowed us to measure and subsequently remove any error from the applied magnetic field (such an error snowballs into an error in the measured atom interferometer phase), and two, that there was no appreciable error to remove, the contribution to the measured atomic phase being well below the 1 mrad level (that is the state-of-the-art at the time of this tutorial), nearing the few $\mu$rad level.
This is entirely expected, considering the electromagnet characteristics with the scaling of current noise to noise in the applied field and finally the transfer of that the error induced on measuring occupation density in the magnetically insensitive Zeeman sublevels (m$_{\text{F}} = 0$ states). 
\section*{Conclusions}
The single chip current sources presented are shown to be capable of consistently producing 3 $\mu$K $^{87}$Rb ensembles when used to control stray magnetic fields and correct field biases as part of red-detuned $\sigma^{+}-\sigma^{-}$ polarization gradient cooling.
We further show these devices are applicable to precision spectroscopy via their application to two-photon Raman spectroscopy, demonstrating no inhomogenous broaden and so an applied field where the transitions are not smeared out to be larger than the applied Rabi frequency. 
This tutorial provides a quick reference for how, with what, and why students and postdocs on cold atom experiments should design, build, and operate these simple circuits to optimize their laser cooling as well as maintain a level of knowledge where tabletop physics directly meets electronics. 
\begin{acknowledgments}
D.O.S. would like to thank all the combative graduate students that made it obvious to those concerned, those being almost exclusively the author, that a tutorial such as this could be relevant and necessary. 
The author would like to thank Valerijus "Val" Gerulis for his electrical engineering support and suggestions, as well as Joseph Junca, Xinhao Zou, and Paul Robert for being patient and studious PhD candidates.
\end{acknowledgments}

\section*{Data Availability Statement}

The data that support the findings of this study are available from the corresponding author upon reasonable request. 
This said, it would be faster to make your own! 
\section*{Conflict of interest}
The authors have no conflicts to disclose, other than those previously stated with various students (who shall remain unnamed for posterity).
\bibliography{coil.bib}

\end{document}